\documentclass[twocolumn,showpacs,showkeys,preprintnumbers,amsmath,amssymb]{revtex4}
\usepackage{graphicx}
\usepackage{epsfig}

\newcommand{\beq}{\begin{equation}}
\newcommand{\eeq}{\end{equation}}
\newcommand{\bea}{\begin{eqnarray}}
\newcommand{\eea}{\end{eqnarray}}
\newcommand{\bwd}{\begin{widetext}}
\newcommand{\ewd}{\end{widetext}}

\begin{document}

\preprint{SLAC-PUB-9628}
\preprint{January 2003}

\title{Calculation of the Coherent Synchrotron Radiation Impedance from a
Wiggler\footnote{Work supported by the Department of Energy contract
DE-AC03-76SF00515}}

\author{Juhao Wu}
\email{jhwu@SLAC.Stanford.EDU}
\author{Tor Raubenheimer}
\email{tor@SLAC.Stanford.EDU}
\author{Gennady Stupakov}
\email{stupakov@SLAC.Stanford.EDU}

\affiliation{Stanford Linear Accelerator Center, Stanford University, Stanford,
CA 94309}

\date{\today 
\\ ,,Submitted to Physical Review Special Topics---Accelerators and Beams}

\begin{abstract}
Most studies of Coherent Synchrotron Radiation (CSR) have only considered the
radiation from independent dipole magnets. However, in the damping rings of
future linear colliders, a large fraction of the radiation power will be
emitted in damping wigglers. In this paper, the longitudinal wakefield and
impedance due to CSR in a wiggler are derived in the limit of a large wiggler
parameter $K$. After an appropriate scaling, the results can be expressed in
terms of universal functions, which are independent of $K$. Analytical 
asymptotic results are obtained for the wakefield in the limit of large and 
small distances, and for the impedance in the limit of small and high 
frequencies.
\end{abstract}



\maketitle

\section{Introduction}
Many modern advanced accelerator projects \cite{NLC01,TESLA01,LCLS02} call for 
short bunches with low emittance and high peak current where coherent 
synchrotron radiation (CSR) effects may play an important role. CSR is emitted 
at wavelengths longer than or comparable to the bunch length whenever the beam 
is deflected \cite{Warnock90,KYNg90}.  The stringent beam requirements needed 
for short wavelength SASE Free-Electron Lasers have led to intensive 
theoretical and experimental studies 
\cite{CARTor95,Mur97,Derb95,Derb96,DL97,SSY97,RLI99,BraunPRL,BraunPRST,HSK02,HK02,SSY02} 
over the past a few years where the focus has been on the magnetic bunch 
compressors required to obtain the high peak currents.  In addition to these 
single-pass cases, it is also possible that CSR might cause a microwave-like 
beam instability in storage rings.  A theory of such an instability in a 
storage ring has been recently proposed in Ref. \cite{SH02} with experimental 
evidence published in \cite{Byrd02}.  Other experimental observations 
\cite{And00,Abo00,Arp01,Car01,Pod01} may also be associated with a CSR-driven 
instability as supported by additional theoretical studies 
\cite{JMW98,MW01,VW02}.

The previous study of the CSR induced instability assumed that the impedance is
generated by the synchrotron radiation of the beam in the storage ring bending 
magnets \cite{SH02}. In some cases (e.g. the NLC damping ring \cite{Andy02}), a
ring will include magnetic wigglers which introduce an additional contribution 
to the radiation impedance. The analysis of the microwave instability in such a
ring requires knowledge of the impedance of the synchrotron radiation in the 
wiggler. Although there have been earlier studies of the coherent radiation
from a wiggler or undulator \cite{YHC90,Sal98}, the results of these papers 
cannot be used directly for the stability analysis.

In this paper, we derive the CSR wake and impedance for a wiggler.  We focus 
our attention on the limit of a large wiggler parameter $K$ because this is the
most interesting case for practical applications.  It also turns out that, in 
this limit, the results can be expressed in terms of universal functions of a 
single variable after an appropriate normalization.

The paper is organized as follows. In Sec. 2, we write down equations for the 
energy loss of a beam in a wiggler. We then derive the synchrotron radiation 
wakefield in the limit of a large wiggler parameter $K$ in Sec. 3. In Sec. 4, 
we obtain the synchrotron radiation impedance for a wiggler, and in Sec. 5 we
discuss our results.

\section{Energy Loss and Longitudinal Wake in Wiggler}

The longitudinal wake is directly related to the rate of energy
loss ${d\mathcal{E}}/{dt}$ of an electron in the beam propagating
in a wiggler. For a planar wiggler, a general expression for
${d\mathcal{E}}/{dt}$ as a function of the position $s$ of the electron
in the bunch and the coordinate $z$ in the wiggler was derived in Ref.
\cite{Sal98}. We reproduce here the results of that work using the
authors' notation:
    \beq
    \frac{d\,\mathcal{E}}{c\,dt}=e^2\,k_w\,\int^s_{-\infty}d\,s^{\prime}\,D(
    \hat{s}-\hat{s}^{\prime},K,\hat{z})\frac{d\,\lambda(s^{\prime})}
    {ds^{\prime}},
    \eeq
where $\lambda(s)$ is the bunch linear density,
    \bea\label{eq2}
    &&D(\hat{s},K,\hat{z})=\frac1{\hat{s}}-2\times
    \nonumber \\
    &&\frac{\Delta-K^2\,B(\Delta,\hat{z})\,
    [\sin\Delta\cos\hat{z}+(1-\cos\Delta)\,\sin\hat{z}]}{\Delta^2+K^2\,B^2(
    \Delta,\hat{z})}\;,
    \eea
    \bea\label{bfunction}
    B(\Delta,\hat{z})
    &=&
    (1-\cos\Delta-\Delta\sin\Delta)\cos\hat{z}
    \nonumber \\
    &+&(\Delta\cos\Delta-\sin\Delta)\sin\hat{z}\;,
    \eea
and $\Delta$ is the solution of the transcendental equation
    \bea\label{s_hat}
    \hat{s}
    &=&
    \frac{\Delta}2\left(1+\frac{K^2}2\right)+\frac{K^2}{4\,\Delta}\{[2(1
    -\cos\Delta)-\Delta\sin\Delta]
    \nonumber\\
    &\times&
    (\cos\Delta\cos\,2\hat{z}+\sin\Delta\sin\,2\hat{z})-2(1-\cos\Delta)\}.
    \eea
In the above equations, we use the following dimensionless variables: 
$\hat{s}=\gamma^2\,k_w\,s$ and $\hat{z}=k_w\,z$. The parameter $\Delta$ is 
equal to $k_w\,(z-z_{r})$, where $z$ and $z_{r}$ are the projected coordinates 
on the wiggler axis of the current position of the test particle and the 
retarded position of the source particle, respectively. The internal coordinate
$s$ is defined so that the bunch head corresponds to a larger value of $s$ than
the tail. The wiggler parameter $K$ is approximately $K\approx 93.4\,B_w\,
\lambda_w$, with $B_w$ the peak magnetic field of the wiggler in units of Tesla
and $\lambda_w$ the period in meters. In addition, $\gamma$ is the Lorentz 
factor, $e$ is the electron charge, $c$ is the speed of light in vacuum, and 
$k_w=2\pi/\lambda_w$ is the wiggler wavenumber. Note that the function $D$ is a
periodic function of $\hat{z}$ with a period equal to $\pi$. Also note that, 
desipte assuming $K \gg 1$, we still assume a small-angle orbit approximation, 
i.e., $K/\gamma \ll 1$.

We introduce the longitudinal wake $W(s)$ of the bunch as the rate of the
energy change averaged over the $z$ coordinate:
    \beq
    W(s)
    =
    -\frac{1}{e^2}
    \frac{d\,\bar{\mathcal{E}}}{c\,dt}
    =
    -k_w\,\int^s_{-\infty}d\,s^{\prime}\,
    G({s}-{s}^{\prime})\frac{d\,\lambda(s^{\prime})}{ds^{\prime}},
    \eeq
where
    \beq\label{Gfunction}
    G(s)
    =
    \frac1{\pi}\int^{\pi}_0\,d\,\hat{z}\,D(\hat{s},K,\hat{z}),
    \eeq
and we dropped $K$ from the list of arguments of the function $G$.
The positive values of $W$ correspond to the energy loss and the
negative values imply the energy gain. The usual longitudinal wake
$w(s)$ corresponding to the interaction of two particles is then
defined as
    \beq\label{WakeGreen}
    w(s)
    =
    -k_w\,\frac {d\,G(s)}{d\,s}\,\,,
    \eeq
so that
    \beq\label{Ww}
    W(s)
    =
    \int^s_{-\infty}
    ds^{\prime}\,
    w({s}-{s}^{\prime})
    \lambda(s^{\prime})
    .
    \eeq
Note that the wake Eq. (\ref{WakeGreen}) is localized in front of
the particle and vanishes behind it, $w=0$ for $s<0$.

In the limit of large $K$, we can neglect unity in the first
bracket of Eq. (\ref{s_hat}), assuming that $K^2/2\gg 1$. Such an approximation
is valid, if we are not interested in the very short distances of order of
$(K\,k_w\,\gamma^2)^{-1}$ (0.5\AA\, for the NLC damping ring wiggler
\cite{Andy02}). We also  introduce a new variable $\zeta\equiv{\hat{s}}/K^2$
which eliminates the parameter $K$ from Eq. (\ref{s_hat}):
    \bea\label{zeta}
    \zeta(\Delta,\hat{z})
    &=&\frac{\Delta}4+\frac 1{4\Delta}\{[2(1-\cos\Delta)-\Delta
    \sin\Delta]
    \nonumber\\
    &\times&
    (\cos\Delta \cos 2\hat{z}+\sin \Delta \sin 2\hat{z})
    \nonumber \\
    &-&2(1-\cos\Delta)\}.
    \eea
In this limit, the expression for $D$, Eq. (\ref{eq2}) can also be
simplified:
    \beq\label{D_function}
    D(\zeta,\hat{z})
    =
    2\,
    \frac{\sin\Delta\cos\hat{z}+(1-\cos\Delta)\sin\hat{z}}
    {B(\Delta,\hat{z})}\;,
    \eeq
as long as $\Delta$ is not too small, $\Delta\gg1/K$. Again, the parameter
$K$ is eliminated from this equation. A detailed analysis supporting this
approximation can be found in Appendix \ref{OrderOfMagnitude}.

\section{Wakefield}

Using Eq. (\ref{Gfunction}) and (\ref{D_function}) we find
    \beq\label{GLargeK}
    G(\zeta)
    =
    \frac2{\pi}\int^{\pi}_0\,d\,\hat{z}\,\frac{\sin\Delta\cos\hat{z}+(1-
    \cos\Delta)\sin\hat{z}}{B(\Delta,\hat{z})}\;,
    \eeq
where $\Delta=\Delta(\zeta,\hat{z})$ is implicitly determined by
Eq. (\ref{zeta}). The integrand in this equation has singularities
at points $\hat{z}=\hat{z}_s$ where $B (\Delta (\zeta,
\hat{z}_s) , \hat{z}_s) = 0$. It is shown in Appendix
\ref{Scaling} that in the vicinity of a singular point $B( \Delta(
\zeta, \hat{z})) \propto (\hat{z}-\hat{z}_{s})^{1/3}$, and the singularity is
integrable.

\begin{figure}[htbp]
    \begin{center}
    \hspace*{-5mm}\mbox{\epsfig{file=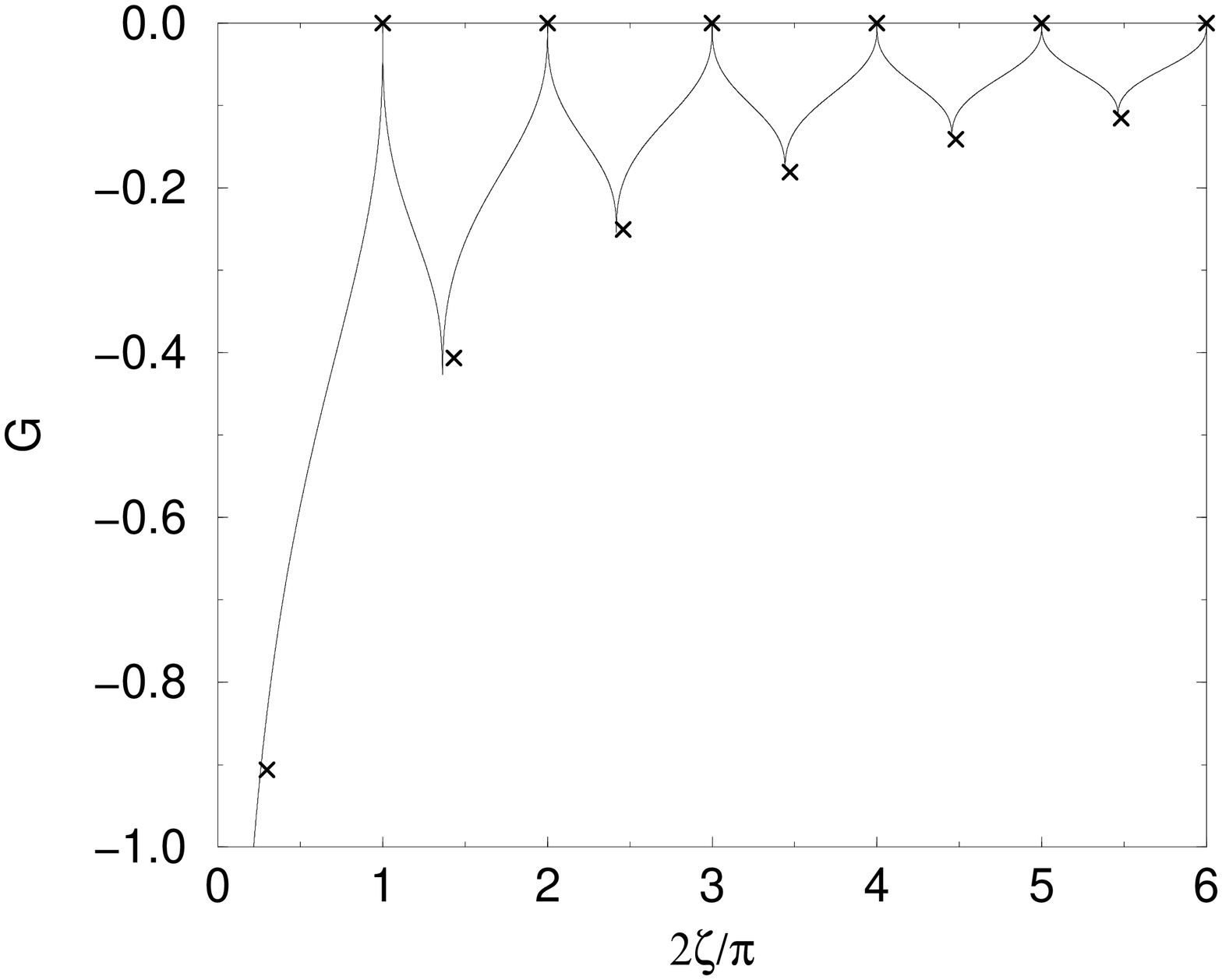, height=9.0cm, angle=-90}}
    \end{center}
    \caption{\label{Gplot}The solid curve represents the $G(\zeta)$ defined in 
    Eq. (\ref{GLargeK}) as a function of the normalized coordinate
    $2\zeta/\pi$. The ($\times$)-signs are the approximation given in Eq. 
    (\ref{GExtreme}).
    }
    \end{figure}

We plot in Fig. \ref{Gplot} the function $G(\zeta)$ calculated by numerical 
integration. A characteristic feature of the function $G$ is the presence of 
cusp points, at which the function reaches local maxima and minima.
An approximate location of these cusp points and the value of the
function $G$ at these points can be understood with a simple
physical argument presented in Appendix \ref{PhysMdl}. It turns
out that the minima are located at distances $s$ between the particles
equal to the integer number of the fundamental radiation wavelength in the
wiggler, and the maxima approximately correspond to the distance
equal to an odd number of half-wavelength. A simple analytical calculation in 
Appendix \ref{PhysMdl} gives the following results
    \bwd
    \bea\label{GExtreme}
    G(\zeta)=\left\{
    \begin{array}{ll}
    0\,\,; & \hspace {0.5cm} \mbox{for } \zeta=\frac{n\,\pi}2 \mbox{ with }
    n=1,2,\cdots \\
    -\frac{4\,(2\,n+1)\,\pi}{4+[(2\,n+1)\,\pi]^2}\,\,;& \hspace
    {0.5cm} \mbox{for } \zeta\approx\frac{(2n+1)\,\pi}4-\frac1{(2n+1)\,\pi}
    \mbox{ with }n=0,1,\cdots
    \end{array}
    \right.
    \eea
    \ewd
These are the ``$\times$'' points in Fig. \ref{Gplot}, showing very good 
agreement with the numerical result.

\subsection{Short-range limit}

In the limit $\zeta \ll 1$, it follows from  Eq. (\ref{zeta}) that
$\Delta \ll 1$ as well. Eq. (\ref{zeta}) can then be solved using
a Taylor expansion of the right-hand side:
    \beq\label{DeltaLeading}
    \Delta=\left(\frac{24\,\zeta}{\cos^2\hat{z}}\right)^{1/3}\;.
    \eeq
Expanding the integrand in Eq. (\ref{GLargeK}), keeping only the
first non-vanishing term in $\Delta$ and substituting $\Delta$
from Eq. (\ref{DeltaLeading}) yields
    \bea\label{GShort}
    G(\zeta)
    &=&
    -\frac1{\pi}\frac{2}{(3\,\zeta)^{1/3}}\int^{\pi}_0d\hat{z}\cos^{2/3}
    \hat{z}
    \nonumber \\
    &=&-\frac{4\,3^{2/3}\,\Gamma\left(\frac{11}6\right)}{5\,
    \sqrt{\pi}\,\Gamma\left(\frac 43\right)}\,\zeta^{-1/3}\approx
    -0.99\,\zeta^{-1/3}.
    \eea
The above result can also be obtained if one considers a wiggler as a 
sequence of bending magnets with the bending radius $R=\gamma/{k_wK|\cos
\hat{z}|}$. Indeed, in a bending dipole, the corresponding ${G_\mathrm{bend}}
(s)=-2\,s^{-1/3} /(3\,R^2)^{1/3}$ \cite{Mur97,Derb95}. Averaging 
${G_\mathrm{bend}}$ over the wiggler period yields Eq. (\ref{GShort}). The 
reason why such a model gives the correct result in this limit, is that the 
formation length of the radiation is much shorter than the wiggler period,
and one can use a local approximation of the bending magnet for the wake.

\subsection{Long-range limit}

In the limit $\zeta \gg 1$, the parameter $\Delta$ is also large, and Eq.
(\ref{zeta}) can be further simplified:
    \beq\label{zetalong}
    \zeta=\frac{\Delta}4-\frac {\sin\Delta\,\cos(\Delta
    -2\hat{z})}{4}\;.
    \eeq

In Eq. (\ref{bfunction}), we keep only the largest term
    \beq
    B(\Delta,\hat{z})=-\Delta\,\sin(\Delta-\hat{z})\;.
    \eeq

For $D$, one now finds,
    \beq
    D(\zeta,\hat{z})\equiv\frac {F(\zeta,\hat{z})}{\zeta}\;,
    \eeq
with
    \beq
    F(\zeta,\hat{z})\equiv\frac{\sin\hat{z}}{2\,\sin(\hat{z}-\Delta(\zeta,
    \hat{z}))}-\frac 12\;,
    \eeq
where the function $\Delta(\zeta,\hat{z})$ is implicitly determined by Eq. 
(\ref{zetalong}). Averaging over one wiggler period, we find
    \beq\label{GLong}
    G(\zeta)\equiv\frac{\bar{F}(\zeta)}{\zeta}\;,
    \eeq
with
    \bea\label{Ffunction}
    \bar{F}(\zeta)
    &\equiv&
    \frac 1{\pi}\,\int^{
    \pi}_0d\,\hat{z}\,F(\zeta,\hat{z})
    \nonumber \\
    &=&
    \frac1{2\,\pi}\left(-\pi+\int_0^{\pi}d\,\hat{z}\,
    \frac{\sin\hat{z}}{\sin(\hat{z}-\Delta)}\right)\;.
    \eea

It is easy to check that the function $\bar{F}$ is periodic, $\bar{F} (\zeta + 
{\pi}/2) = \bar{F}(\zeta)$, and $\bar{F}(0) =0$,  $\bar{F}(\pi/4)=-1$ in 
agreement with Eq. (\ref{GExtreme}). The average value $\bar{F}(\zeta)$ is 
equal to $-1/2$. Since $\bar{F}$ is periodic in $\zeta$ with a period of 
$\pi/2$, using Eq. (\ref{Ffunction}), we get a Fourier series representation 
for $\bar{F}(\zeta)$:

    \bwd
    \beq\label{F_Fourier}
    \bar{F}(\zeta)
    =-\frac12
    +\frac12\sum^{\infty}_{n=0}\left[J_n
    \left(\frac{2n+1}2\right)-J_{n+1}\left(\frac{2n+1}2\right)\right]^2
    \cos(4(2n+1)\zeta)\;,
    \eeq
    \ewd

where $J_n(x)$ is the Bessel function of the first kind. Derivations of the 
Fourier coefficients are presented in Appendix \ref{FourierCoef}. In Fig. 
\ref{FPlot}, we plot the function $\bar{F}(\zeta)$ defined in Eq. 
(\ref{F_Fourier}) for one period.

The corresponding long-range wake is then

    \bwd
    \beq\label{G_Fourier}
    G(\zeta)
    =-\frac1{2\zeta}
    +\frac1{2\zeta}\sum^{\infty}_{n=0}\left[J_n
    \left(\frac{2n+1}2\right)-J_{n+1}\left(\frac{2n+1}2\right)\right]^2
    \cos(4(2n+1)\zeta)\;.
    \eeq
    \ewd

It is worth noting, that the asymptotic expression in the limit $\zeta \gg 1$  
in Ref. \cite{Sal98} is incorrect---instead of the $\bar{F}$-function
the authors obtained a sine function, which only corresponds to
the fundamental mode of the radiation and neglects contribution
from higher-order harmonics.

\begin{figure}[htbp]
\begin{center}
\hspace*{-5mm}\mbox{\epsfig{file=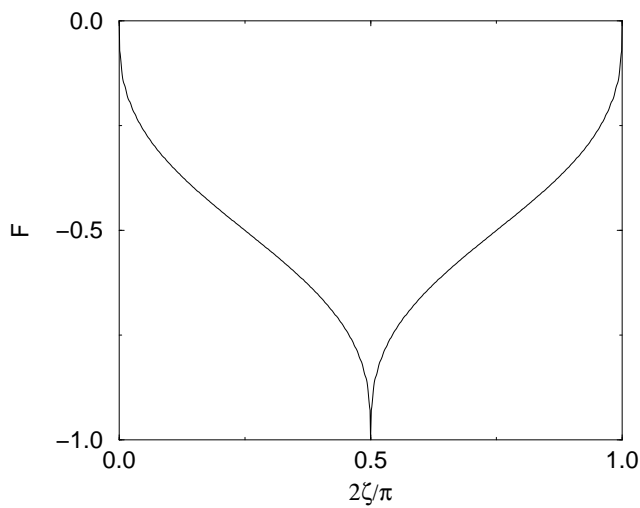, height=9.0cm, angle=-90}}
\end{center}
\caption{\label{FPlot}Plot of $\bar{F}(\zeta)$ of Eq. (\ref{F_Fourier}).
}
\end{figure}

\begin{figure}[htbp]
\begin{center}
\mbox{\epsfig{file=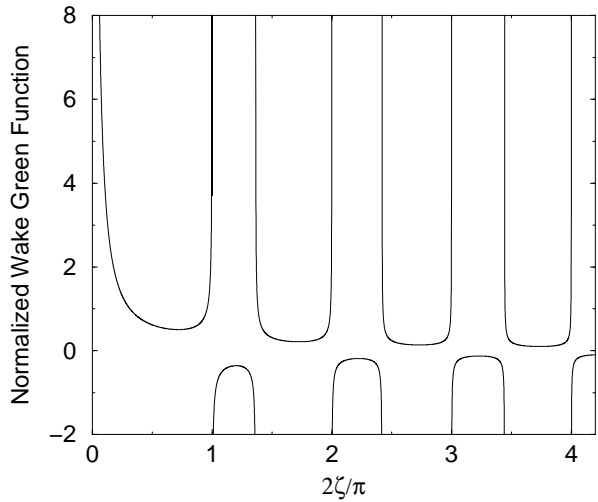, height=9.0cm, angle=-90}}
\end{center}
\caption{\label{RegWakePlot}The normalized wake Green function $-w(s)[K/(\gamma
k_w)]^2$ as a function of the normalized coordinate $2\zeta/\pi$.
}
\end{figure}

The longitudinal wake defined in Eq. (\ref{WakeGreen}) is plotted in Fig.
\ref{RegWakePlot}.

\section{Impedance}

    \begin{figure}[htbp]
    \begin{center}
    \hspace*{-5mm}\mbox{\epsfig{file=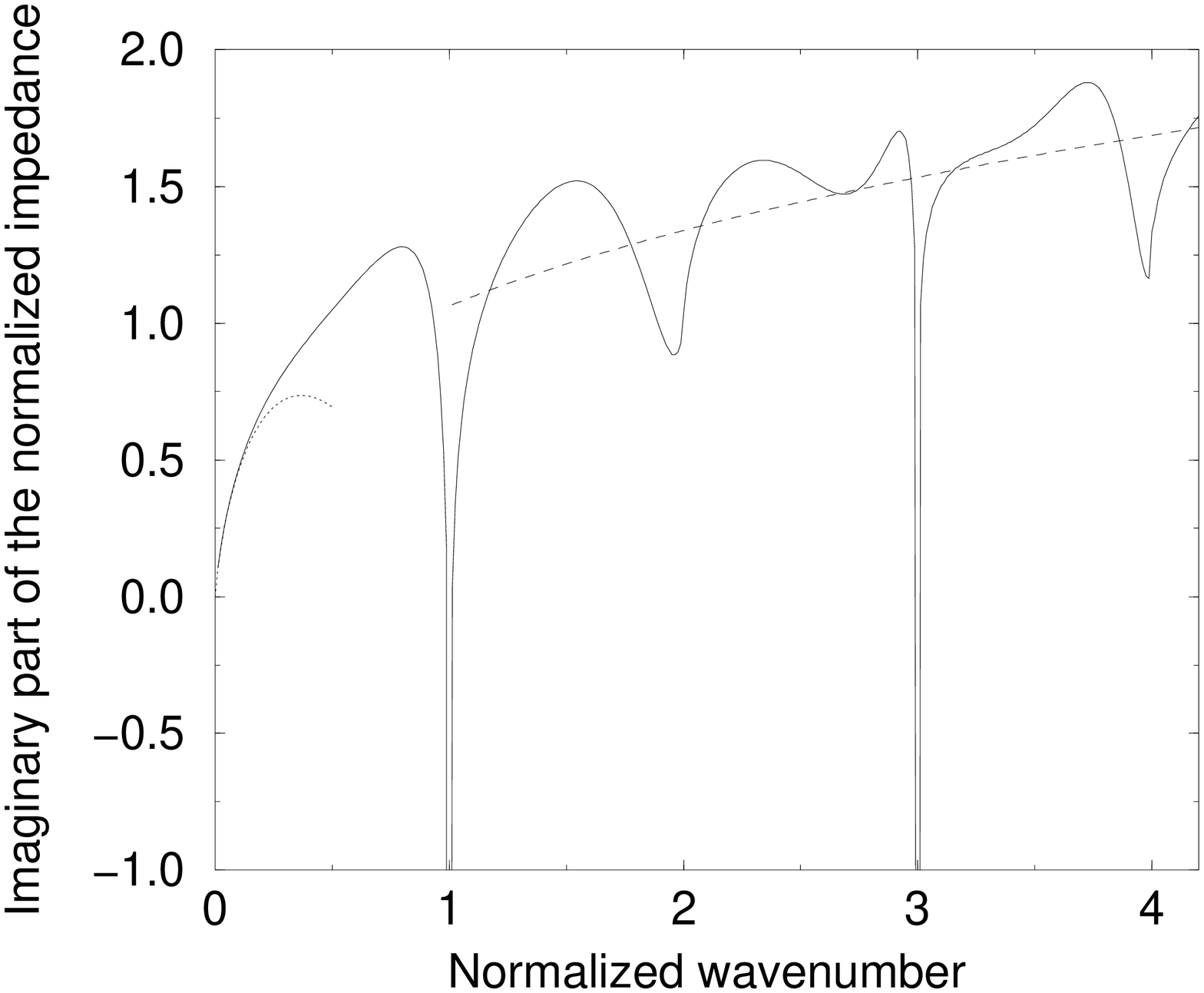, height=9.0cm, angle=-90}}
    \end{center}
    \caption{\label{ImpImg}The imaginary part of the normalized impedance
    $Z(k)/k_w$ as a function of the normalized wavenumber $k/{k_0}$.
    Solid line---numerical solution from Eq. (\ref{impedance_eq}), dotted 
    line---analytical low-frequency asymptotic behavior from Eq. 
    (\ref{ZLowExact}), and dashed 
    line---analytical high-frequency asymptotic behavior from Eq. 
    (\ref{ZHigh}).
    }
    \end{figure}

    \begin{figure}[htbp]
    \begin{center}
    \hspace*{-5mm}\mbox{\epsfig{file=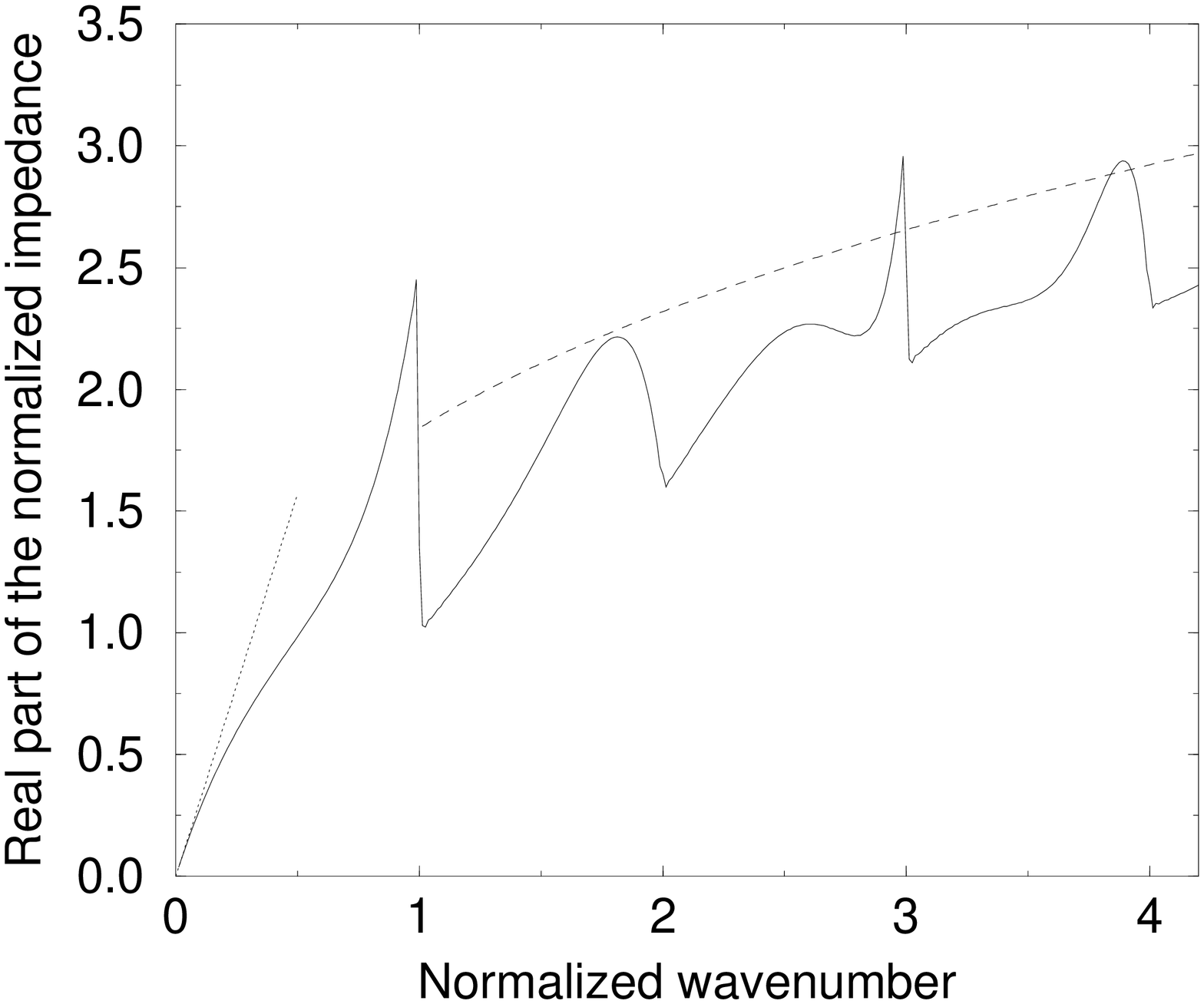, height=9.0cm, angle=-90}}
    \end{center}
    \caption{\label{ImpRel}The real part of the normalized impedance
    $Z(k)/k_w$ as a function of the normalized wavenumber $k/{k_0}$.
    Solid line---numerical solution from Eq. (\ref{impedance_eq}), dotted 
    line---analytical low-frequency asymptotic behavior from Eq. 
    (\ref{ZLowExact}), and dashed 
    line---analytical high-frequency asymptotic behavior from Eq. 
    (\ref{ZHigh}).
    }
    \end{figure}

The impedance $Z(k)$ is defined as the Fourier transform of the
wake,
    \bea\label{impedance_eq}
    Z(k)
    &=&
    \int^{\infty}_0\,ds\,w(s)\,e^{-iks}
    \nonumber \\
    &=&
    -\,i\,k\frac{K^2}{\gamma^2}\int^{\infty}_0\,d\zeta\,G(\zeta)\,e^{-\,4\,i
    \frac k{k_0}\zeta}\;,
    \eea
where $k_0\equiv4\gamma^2k_w/K^2$ is the wiggler fundamental radiation
wavenumber.

We evaluated the integral in Eq. (\ref{impedance_eq}) using
numerically calculated values of the function $G(\zeta)$ in the
interval $[\zeta_{\min},\zeta_{\max}]$, where $\zeta_{\min}\approx
10^{-3}$ and $\zeta_{\max}\approx 50$. The contribution to the
integral outside of this interval was calculated using asymptotic
representations Eqs. (\ref{GShort}) and (\ref{G_Fourier}).

The resulting imaginary and real parts of the impedance are shown
in Fig. \ref{ImpImg} and Fig. \ref{ImpRel} respectively.

The real part of the impedance can be related to the wiggler
radiation spectrum $I(\omega)$ \cite{Chao93}:
   \beq\label{ImpSpec}
   \mathrm{Re}Z(\omega)=\frac {\pi}{e^2}\,I(\omega)\,.
   \eeq
The spectrum $I(\omega)$ in the limit $K \gg 1$ is calculated in
Appendix \ref{Spectrum}. It shows a perfect agreement with the
result presented in Fig. \ref{ImpRel}.

Simple analytical formulae for the impedance can be obtained in
the limit of low and high frequencies.

The low-frequency impedance corresponds to the first term in
Eq. (\ref{G_Fourier}) for function $G$ which does not oscillate with $\zeta$:
    \beq\label{GLongAsy}
    G(\zeta)=-\frac 1{2\,\zeta}.
    \eeq
Using the definition in Eq. (\ref{impedance_eq}), we then obtain the
low-frequency asymptotic behavior of the impedance as
    \bea\label{ZLowExact}
    Z(k)
    &=&
    -\,i\,2\,k_w\frac k{k_0}\left[\gamma_{\mathrm{E}}+\log\left(\frac{4k}{k_0}
    \right)+i\frac{\pi}2\right]
    \nonumber \\
    &\approx&\pi\,k_w\frac k{k_0}\left[1
    -\frac{2\,i}{\pi}\,\log\left(\frac k{k_0}\right)\right]\;,
    \eea
where, $\gamma_{\mathrm{E}}\approx0.5772$ is the Euler Gamma constant. This 
asymptotic low-frequency impedance is plotted in Figs. \ref{ImpImg} and 
\ref{ImpRel} for comparison with the numerical solution.

Since we have an analytical expression for the short-range
$G(\zeta)$ in Eq. (\ref{GShort}), we get the asymptotic high-frequency
impedance as
    \bea\label{ZHigh}
    Z(k)
    &=&-i\,\frac{6\,\Gamma\left[\frac {11}6\right]}{5\,\sqrt{\pi}\,\Gamma
    \left[\frac 43\right]}\,A\,\left(\frac{Kk_w}{\gamma}\right)^{2/3}k^{1/3}
    \nonumber \\
    &\approx&-\,0.71\,i\,A\,\left(\frac{Kk_w}{\gamma}\right)^{2/3}k^{1/3}\;,
    \eea
with $A=3^{-1/3}\Gamma(2/3)(\sqrt{3}\,i-1)\approx 1.63\,i-0.94$ \cite{SH02}.
This asymptotic high-frequency impedance is plotted in Figs. \ref{ImpImg} and
\ref{ImpRel} for comparison with the numerical solution.

\section{Discussion and conclusion}

In this paper, we derived the wakefield and the impedance for wigglers with 
${K^2}/2 \gg 1$ due to the synchrotron radiation. Analytical asymptotic results
are obtained for the wakes in the limit of small and large distances, and for 
the impedance in the limit of small and high frequencies. The results obtained 
in this paper are used for the beam instability study due to the synchrotron 
radiation in wigglers \cite{WSRH02}.

\begin{acknowledgments}

The authors thank Drs. A.W. Chao, S.A. Heifets, Z. Huang of Stanford Linear
Accelerator Center, Drs. S. Krinsky, J.B. Murphy, J.M. Wang of National
Synchrotron Light Source, Brookhaven National Laboratory for many discussions.
Work was supported by the U.S. Department of Energy under contract
DE-AC03-76SF00515.

\end{acknowledgments}

\appendix

\section{Details for deriving Eq. (\ref{D_function})}\label{OrderOfMagnitude}
In the limit of $\Delta\ll1$, according to Eq. (\ref{bfunction}), $B\sim
\Delta^2$. In the numerator of the second term on the right hand side of Eq.
(\ref{eq2}), we would have $K^2B(\Delta,\hat{z})[\sin\Delta\cos\hat{z}+(1-\cos
\Delta)\sin\hat{z}]\sim K^2\Delta^3\gg\Delta$, as long as $\Delta\gg1/K$. This
is allowed, since we are interested in the limit of $K\gg1$. Hence, $\Delta$ is
neglected in the numerator. In the denominator of the second term, then
$K^2B^2\sim K^2\Delta^4\gg\Delta^2$, as long as $\Delta\gg 1/K$, hence
$\Delta^2$ is dropped. Therefore, the second term of Eq. (\ref{eq2}) is on the
order of $1/\Delta$. According to Eq. (\ref{s_hat}), in the limit of $\Delta
\ll1$, we have $\hat{s}\sim K^2\Delta$, so the first term of Eq. (\ref{eq2}) is
on the order of $1/(K^2\Delta)$, hence is much smaller than the second term in 
the limit of $K\gg 1$. Therefore the first term $1/\hat{s}$ could be dropped. 
All these considerations lead us to Eq. (\ref{D_function}).

For $\Delta \sim 1$ and $K \gg 1$, according to Eq. (\ref{s_hat}), we have
$\hat{s} \sim  K^2 \gg 1$. Eq. (\ref{bfunction}) suggests that
$B(\Delta,\hat{z}) \sim 1$. Now, in Eq. (\ref{eq2}), in the limit
of $K\gg1$, we can neglect $\Delta$ in comparison with $K^2$ in
the numerator and $\Delta^2$ in the denominator of the second term
on the right hand side. We then note that the second term is on
the order of 1, and is much larger than the first term $1/\hat{s}
\sim K^{-2} \ll 1$, hence we can drop $1/{\hat{s}}$ to obtain Eq.
(\ref{D_function}).

Now let us study the limit of $\Delta\gg1$. Eq. (\ref{bfunction}) suggests that
$B(\Delta,\hat{z}) \sim \Delta$. For $K\gg1$, then in Eq. (\ref{eq2}),
$\Delta$ and $\Delta^2$ could be dropped in the numerator and denominator of
the second term on the right hand side, respectively. The second term is on the
order of $1/\Delta$. Now, according to Eq. (\ref{s_hat}), in the limit of
$\Delta\gg1$, we have $\hat{s}\sim K^2\Delta$. Hence, in the limit of $K\gg1$, 
the first term of Eq. (\ref{eq2}), which is on the order of $1/(K^2\Delta)$, is
negligible, compared with the second term. Hence, we obtained Eq.
(\ref{D_function}).

So, in general, for large $K$, as long as $\Delta$ is not too small, i.e.,
$\Delta\gg 1/K$, the simplification leading to Eq. (\ref{D_function}) is always
acceptable.

\section{Singular points in $D(\zeta,\hat{z})$}\label{Scaling}

To find the scaling of the singularity, we assume that at the vicinity of the
zeroes $\hat{z}_s$ of $B(\Delta,\hat{z})$, the leading term scales as
    \beq\label{B_initial}
    B\approx b\,(\hat{z}-\hat{z}_s)^{\alpha}\,,
    \eeq
then we have
    \beq\label{B_Prime_initial}
    B^{\prime}\approx
    \alpha\,b\,(\hat{z}-\hat{z}_s)^{\alpha-1}\,,
    \eeq
where the prime indicates the
derivative with respect to $\hat{z}$.

Let us first calculate $B'$. From Eq. (\ref{bfunction}) we have,
    \bea\label{B_Prime}
    B^{\prime}
    &=&-\sin\hat{z}+\sin(\hat{z}-\Delta)
    \nonumber \\
    &+&\cos(\hat{z}-\Delta)\Delta
    -\cos(\hat{z}-\Delta)\Delta\Delta^{\prime}\;.
    \eea
To find $\Delta'$, we revert to Eq. (\ref{zeta}), where, we find
    \beq\label{DeltaPrime}
    \Delta^{\prime}=\frac{C(\Delta,\hat{z})}{B^2(\Delta,\hat{z})}
    \eeq
with
    \bea\label{C_Expression}
    C(\Delta, \hat{z})
    &=&2\sin(2\hat{z}-\Delta)\sin\left(\frac{\Delta}2\right)\Delta
    \nonumber \\
    &\times&\left[2\sin\left(\frac{\Delta}2\right)-\cos\left(\frac{
    \Delta}2\right)\Delta\right]\;.
    \eea
Note that $C(\Delta, \hat{z})$ is a well defined function at the zeroes of
$B(\Delta, \hat{z})$.

Combining Eqs. (\ref{B_initial}), (\ref{B_Prime}), (\ref{DeltaPrime}),
(\ref{C_Expression}), we have, near the zeroes $\hat{z}_s$,
    \beq\label{BPrimeCal}
    B^{\prime}(\Delta,\hat{z})=\frac {\Xi}
    {b^2\,(\hat{z}-\hat{z}_s)^{2\alpha}}\;,
    \eeq
where,
    \bwd
    \bea
    \Xi
    &=&-\cos(\hat{z}-\Delta)\Delta\times C(\Delta, \hat{z})|_{\hat{z}=\hat{z}
    _{s};\Delta=\Delta(\zeta,\hat{z}_{s})}
    \nonumber \\
    &=&-\cos(\hat{z}-\Delta)\Delta\left\{2\,\sin(2\hat{z}-\Delta)\sin\left(
    \frac{\Delta}2\right)\Delta
    \times\left.\left[2\sin\left(\frac{\Delta}2\right)-\cos\left(
    \frac{\Delta}2 \right)\Delta\right]\right\}\right|_{\hat{z}=\hat{z}
    _{s};\Delta=\Delta(\zeta,\hat{z}_{s})}\;.
    \eea
    \ewd
Here, $\hat{z}_{s}$ is defined as the solution of $B(\Delta(\zeta,\hat{z}_{s}),
\hat{z}_{s})=0$. Therefore, combining Eqs. (\ref{BPrimeCal}) and 
(\ref{B_Prime_initial}), we find the scaling index $\alpha=1/3$ and $b=(3\,
\Xi)^{1/3}$. This means that $D(\zeta,\hat{z})$ has only an integrable 
singularity at $z=z_s$ with $D \propto (\hat{z}-\hat{z}_{s})^{-1/3}$.

\begin{figure}[htbp]
    \begin{center}
    \hspace*{-5mm}\mbox{\epsfig{file=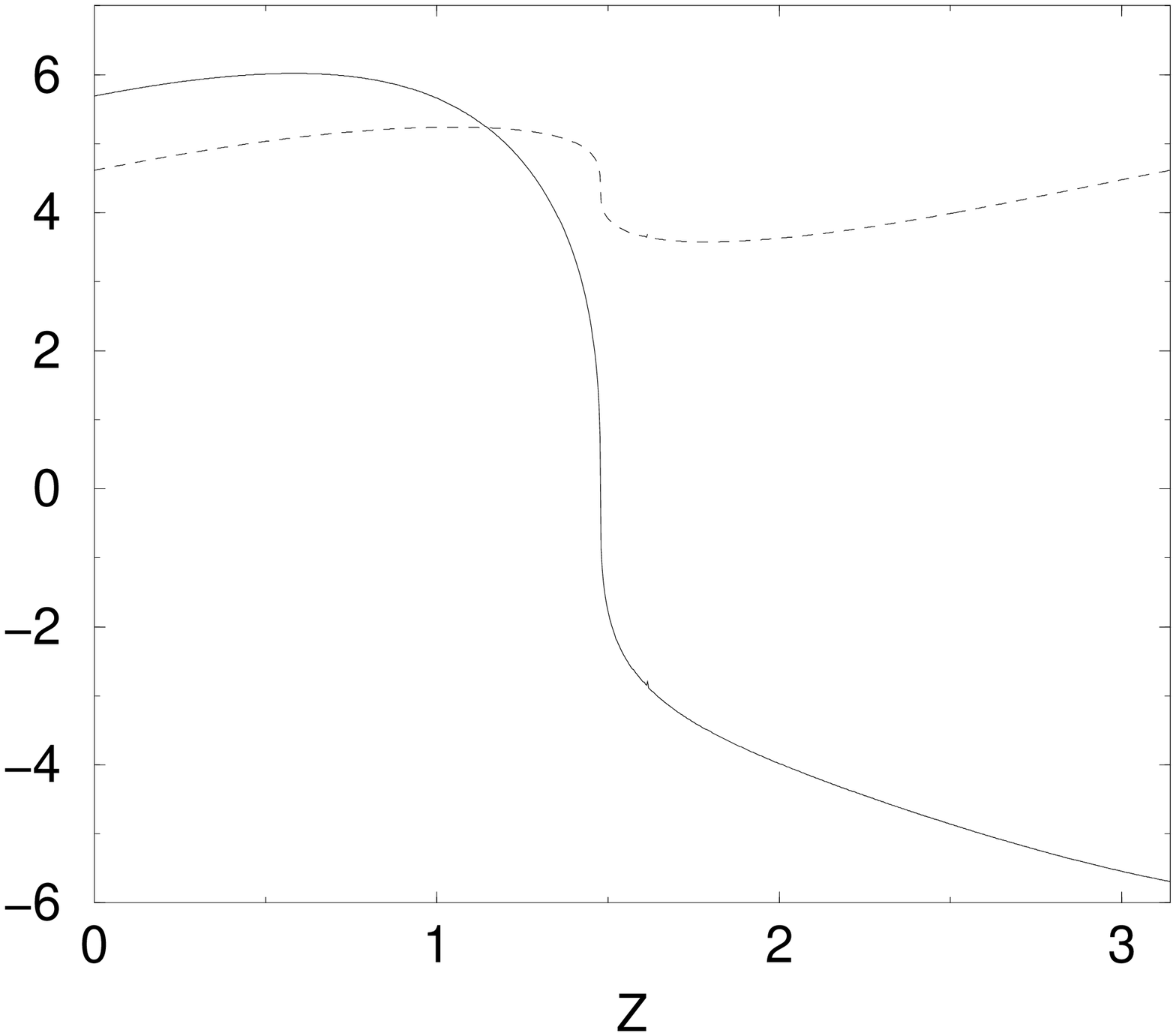, height=9.0cm, angle=-90}
    }
    \end{center}
    \caption{\label{singularity}A concrete example for $\zeta=1.0$. Solid line:
    $B(\Delta(\zeta=1.0,\hat{z}),\hat{z})$ as a function of $\hat{z}$; Dashed
    line: $\Delta(\zeta=1.0,\hat{z})$ as a function of $\hat{z}$.
    }
    \end{figure}

As a numeric illustration of the origin of the singularity, we plot in Fig.
\ref{singularity} functions
$B(\Delta(\zeta,\hat{z}),\hat{z})$ and $\Delta(\zeta,\hat{z})$
for $\zeta=1.0$.
This plot shows that at the point where $B=0$, both derivatives $B'$ and
$\Delta'$ are infinite, in accordance with Eqs. (\ref{B_Prime_initial})
and (\ref{DeltaPrime}).

\section{Simple physics model}\label{PhysMdl}

\begin{figure}[htbp]
\begin{center}
\hspace*{-5mm}\mbox{\epsfig{file=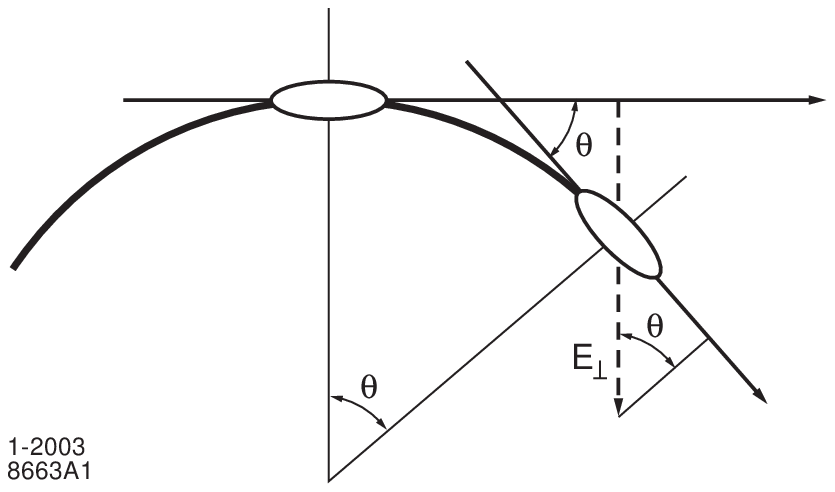, height=5.0cm, angle=0}}
\end{center}
\caption{\label{DerbenevModel}Physics model to explain the longitudinal CSR 
wake.
}
\end{figure}

Here, we give some explanation about the peaks and the zeroes in Fig.
\ref{Gplot} based on a simple physics model.
\begin{figure}[htbp]
\begin{center}
\hspace*{-5mm}\mbox{\epsfig{file=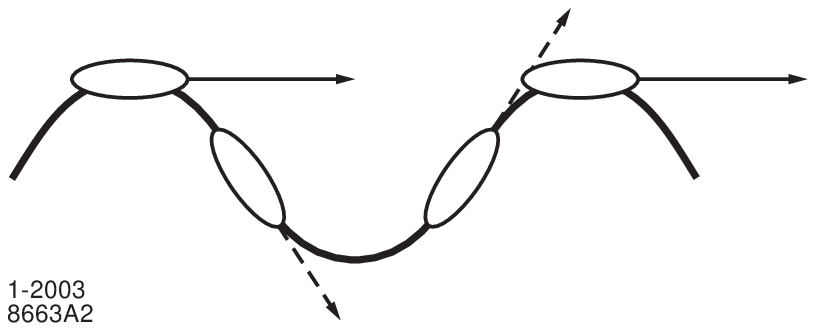, height=3.0cm, angle=0}}
\end{center}
\caption{\label{TorModel}Physics model to explain the peaks and zeroes of
$G(s)$ shown in Fig. \ref{Gplot}. The solid line stands for the electron
trajectory. The ellipses stand for the electrons. The arrows stand for
the instantaneous direction of the motion. We group the four electrons into two
pairs, one with solid arrows and the other with dashed arrows.
}
\end{figure}

The CSR wake is actually the field emitted by a trailing particle which
acts on the particle in front. Using a model presented in Ref. \cite{Derb95}, 
the longitudinal force on the leading particle can be thought of as the 
component of the trailing particle's transverse Coulomb field projected onto 
the leading particle's direction of motion:
    \beq\label{LongF}
    W_{||}=eE_{\perp}(z)\sin\theta\;,
    \eeq
where $E_{\perp}(z)$ is the magnitude of the transverse Coulomb electric field 
at the retarded position from the trailing particle at retarded time. The 
argument $z$ indicates that the amplitude of the transverse electric field is 
actually varying along the trajectory. In Eq. (\ref{LongF}), $\theta$ is the 
angle between the direction of motion of the trailing particle and that of the 
particle at front. To illustrate the model, we give a schematic plot in Fig. 
\ref{DerbenevModel}.

To understand the wiggler wakefield, let us look at the four electrons in 
Fig. \ref{TorModel}. The pair with solid arrows is separated by integer number 
times of the wiggler fundamental radiation wavelength. During their journey, 
when the light emitted by the trailing electron cathches the electron in front,
the instantaneous direction of motion of the front electron is always parallel 
to the direction of motion of the trailing electron at the retarded time when 
it emitted the light. Hence we have $\theta=0$. So according to Eq. 
(\ref{LongF}), the longitudinal force is always zero. This explains the zeroes 
in the longitudinal wake potential plotted in Fig. (\ref{Gplot}). The pair with
dashed arrows is separated by odd integer number times of half of the wiggler 
fundamental radiation wavelength. Averaged over one period, they make the 
largest angle between the instantaneous direction of motion of the front 
electron and the direction of motion of the trailing electron at the retarded 
time. Hence according to Eq. (\ref{LongF}), the longitudinal force reaches 
maximum. This explains the peaks shown in Fig. (\ref{Gplot}).

Let us calculate the values at these cusps points. According to Eq. 
(\ref{zeta}), when $\Delta=2n\pi$, we have $\zeta=n\pi/2$. According to Eq.
(\ref{GLargeK}), the numerator of the integrand is zero at $\Delta=2n\pi$,
while $B(\Delta,\hat{z})=2n\pi\sin\hat{z}$ according to Eq. (\ref{bfunction}).
Hence the integral is zero. So we have $G(n\pi/2)=0$ for $n=1,2,\cdots$. We
find that it is true in the numerical solution in Fig. \ref{Gplot}.

At $\Delta=(2n+1)\pi$, according to Eqs. (\ref{GLargeK}) and (\ref{bfunction}),
we get
    \bea
    &&G(\zeta)
    =
    \frac2{\pi}\int_0^{\pi}d\hat{z}\frac{2\sin\hat{z}}{2\cos
    \hat{z}-(2n+1)\pi\sin\hat{z}}
    \nonumber \\
    &=&-\frac{4\,(2\,n+1)\,\pi}{4+[(2\,n+1)\,\pi]^2}\hspace{0.5 cm}
    {\mbox{for}}\hspace{0.2 cm}\Delta=(2n+1)\,\pi.
    \eea
According to Eq. (\ref{zeta}), we have
    \beq
    \zeta=\frac{\Delta}4-\frac2{\Delta}\,\cos^2\hat{z}\approx\frac{\Delta}4
    -\frac1{\Delta}\hspace{0.5 cm}{\mbox{for}}\hspace{0.2 cm}\Delta=(2n+1)\,
    \pi.
    \eeq
In the above approximation, we average $\zeta$ over one period in $\hat{z}$.
This becomes a good approximation, when $\Delta$ is large. This manifests
itself in Fig. \ref{Gplot}. As we find from Fig. \ref{Gplot}, with the
increasing of $\zeta$, therefore, the increasing of $\Delta$, the above
approximate value gets closer and closer to the numerical solution. Combining
the results at the zeroes and the peaks, we get Eq. (\ref{GExtreme}).

\section{Calculation for the Fourier coefficients}\label{FourierCoef}
Since we find that the function $\bar{F}(\zeta)$ is periodic in $\zeta$ with a
period of $\pi/2$, we could represent it in a Fourier series. The calculation 
for the Fourier coefficients are straightforward. We here illustrate one
example. For $m=1,2,\cdots$ ,
	\bwd
	\bea\label{CosCoef}
	&&
	\langle\bar{F}(\zeta)\cos[4(2m+1)\zeta]\rangle\equiv\frac 4{\pi}
	\int^{\frac{\pi}2}_0d\zeta\bar{F}({\zeta})\cos[4(2m+1)\zeta]
	\nonumber \\
	&=&
	\frac2{\pi^2}\int^{\frac{\pi}2}_0d\zeta\cos[4(2m+1)\zeta]\int^{\pi}_0
	d\hat{z}\frac{\sin{\hat{z}}}{\sin(\hat{z}-\Delta)}
	\nonumber \\
	&=&
	\frac1{\pi^2}\int^{2\pi}_0d\Delta\int^{\pi}_0d\hat{z}\sin\hat{z}
	\sin(\hat{z}-\Delta)\cos\left[(2m+1)\Delta-\frac{(2m+1)\sin[2(\Delta
	-\hat{z})]}2-\frac{(2m+1)\sin2\hat{z}}2\right]
	\;.
	\eea
	\ewd
Notice that, we have used the definition of $\bar{F}$ in Eq. (\ref{Ffunction}).
We also changed integral variable pair $(\zeta,\hat{z})$ to $(\Delta,\hat{z})$,
using Jacobian obtained from Eq. (\ref{zetalong}). To complete the integral in
Eq. (\ref{CosCoef}), we make use of the well-known identities
	\beq
	\cos(z\cos\theta)=\sum^{\infty}_{n=0}\epsilon_{2n}(-)^nJ_{2n}(z)
	\cos2n\theta\;,
	\eeq
with, $\epsilon_0=1$, $\epsilon_n=2$ for $n=1,2,\cdots$ ,
and
	\beq
	\sin(z\cos\theta)=2\sum^{\infty}_{n=0}(-)^nJ_{2n+1}(z)
	\cos(2n+1)\theta\;.
	\eeq
All the other Fourier coefficients, including the average value $\langle
\bar{F}(\zeta)\rangle=-1/2$, are obtained in the same manner. We then obtain
the Fourier series representation in Eq. (\ref{F_Fourier}).

\section{Real part of the impedance and the wiggler radiation
spectrum}\label{Spectrum}

Our approach to calculate $I(\omega)$ is based on the paper of Alferov {\it{et
al.}} \cite{Alf73} and that of Krinsky {\it{et al.}} \cite{Kri83}. The only
difference is that we are dealing with large $K$ case, hence we could eliminate
$K$ from the equations as was done for the impedance. Let us illustrate this
in the following. The energy radiated per electron per unit solid angle per 
unit frequency interval per unit length is given by
    \beq
    \frac{dI(\omega)}{d\Omega}=\frac{e^2\,\gamma^2\,N}
    {c\,\lambda_w}\sum^{\infty}_{m=1}G_{m}(K,\gamma\,\theta,\phi)
    H_{m}\left(\frac{\omega}{\omega_1}\right)\;,
    \eeq
where $m$ is the harmonic number of frequencies in the spectrum. $N$ is the
number of periods of the wiggler. The polar coordinates $\theta$ and $\phi$
are defined so that, $\theta=0$ corresponding to the forward direction along
the wiggler axis, and $\phi=0$ to be the plane of electron motion. For a
large $K$, $\omega_1 \approx{2\,c\,k_w\,\gamma^2}/({K^2/2+\gamma^2\,\theta^2})$
is the fundamental radiation frequency, and
    \bea
    G_{m}(K,\gamma\,\theta,\phi)&\approx&\frac{4\,m^2}{\left(\frac12\,
    K^2+\gamma^2\,\theta^2\right)^2}\left\{\left[S_1\,\gamma\,\theta\,
    \cos\phi\right.\right.
    \nonumber \\
    &-&
    \left.\left.\left(S_1+\frac2{m}\,S_2\right)
    \frac{\frac12\,K^2+\gamma^2\,\theta^2}{2\,\gamma\,\theta\,\cos\phi}
    \right]^2\right.
    \nonumber \\
    &+&
    \left.(\gamma\,\theta)^2\,S_1^2\,\sin^2\phi\right\}\;,
    \eea
where, we define
    \beq
    \xi_z\equiv\frac{K^2}{{4\left(\frac{K^2}2+\gamma^2\,\theta^2\right)}}\;,
    \eeq
and
    \beq
    \xi_x\equiv\frac{2\,K\,\gamma\,\theta\,\cos\phi}{\frac{K^2}2+\gamma^2\,
    \theta^2}\;,
    \eeq
and in turn
    \beq
    S_1\equiv\sum^{\infty}_{n=-\infty}J_n(m\,\xi_z)\,
    J_{2n+m}(m\,\xi_x)\;,
    \eeq
and
    \beq
    S_2\equiv\sum^{\infty}_{n=-\infty}n\,J_n(m\,\xi_z)\,J_{2n+m}(m\,
    \xi_x)\;.
    \eeq

For finite, but large $N$,
    \beq
    H_{m}\left(\frac{\omega}{\omega_1}\right)=\frac{\sin^2\left[N\,\pi\,\left(
    \frac{\omega}{\omega_1}-m\right)\right]}
    {\pi^2\,N^2\,\left(\frac{\omega}{\omega_1}-m\right)^2}\;,
    \eeq
determines the bandwidth of the radiation. In our theory, we focus on an
infinite long wiggler, i.e., we are interested in the limit of $N\rightarrow
\infty$. Under such condition, we have $H_{m}\rightarrow{\omega_1}\,
\delta(\omega-m\,\omega_1)/N$. Now we are ready to get the radiation spectrum
by integrating over the entire solid angle.

    \beq\label{spectrum}
    I(\omega)=\int^{\pi}_0d\,\theta\,\sin\theta\int^{2\pi}_0d\,\phi\frac{
    d\,I(\omega)}{d\,\Omega}.
    \eeq

Due to the $\delta$-function in $H_m$, the integral over $\theta$ could be done
first. Also due to the fact that $\theta$ is small, we approximate
$\sin\theta\approx\theta$. Then the integral in Eq. (\ref{spectrum}) is reduced
into a 1-D integral as the following,
    \bea\label{spec_rev}
    I(k)&=&\frac{4\,e^2}{c\,\lambda_w}\sum^{\infty}_{m=1}
    \int^{\frac{\pi}2}_{-\frac{\pi}2}\,d\,\phi\,\left(m-\frac k{k_0}\right)
    \nonumber \\
    &\times&
    \left\{\left[S_1\cos\phi-\left(S_1+\frac 2{m}\,S_2\right)\frac {m}
    {2\,\cos\phi\,\left(m-\frac k{k_0}\right)}\right]^2\right.
    \nonumber \\
    &+&
    \left.S_1^2\,\sin^2\phi\right\}\;.
    \eea

Notice, the integrand has a period of $\pi$, hence we need only integrate for
one period. $S_1$ and $S_2$ are further simplified as
    \bea
    S_1&=&\sum^{\infty}_{n=-\infty}J_n\left(\frac k{2k_0}\right)
    \nonumber \\
    &\times&J_{2n+m}\left(
    \cos\phi\,\sqrt{\frac{8k}{k_0}\,\left(m-\frac k{k_0}\right)}\right)\;,
    \eea
and
    \bea
    S_2&=&\sum^{\infty}_{n=-\infty}n\,J_n\left(\frac k{2k_0}\right)
    \nonumber \\
    &\times&
    J_{2n+m}\left(\cos\phi\,\sqrt{\frac {8k}
    {k_0}\,\left(m-\frac k{k_0}\right)}\right)\;.
    \eea

Now based on Eqs. (\ref{ImpSpec}) and (\ref{spec_rev}), we could compute the
real part of the impedance as
    \bwd
    \beq
    \mbox{Re}[Z(k)]\equiv 2\,k_w\,\sum^{\infty}
    _{m=1}\int^{\frac{\pi}2}_{-\frac{\pi}2}\,d\,\phi\,\left(m-\frac k{k_0}
    \right)
    \times
    \left\{\left[S_1\cos\phi-\left(S_1+\frac 2{m}\,S_2\right)\frac {m}
    {2\,\cos\phi\,\left(m-\frac k{k_0}\right)}\right]^2+S_1^2\,\sin^2\phi
    \right\}\;.
    \eeq
    \ewd

Notice $K$ is eliminated from this equation, as long as $K$ is large. We sum up
the first 30 harmonics to obtain the real part of the normalized impedance up 
to the 4th harmonic. Adding higher harmonics will not change the impedance 
within this frequency region within numerical accuracy. The result is identical
to Fig. \ref{ImpRel}, supporting the validity of our calculation.

\end{document}